\documentclass[11pt,twoside]{article}
\usepackage{graphicx,epsfig,natbib,epstopdf}
\usepackage{CS18}
\begin{document}
%
%
%
\title{Zeeman Doppler imaging of a cool star using line profiles in all four Stokes parameters for the first time}
%
%
\author{L.~Ros\'{e}n$^1$, O.~Kochukhov$^1$, G.~A.~Wade$^2$}
\affil{$^1$ Department of Physics and Astronomy, Uppsala University, Box 516, 751 20 Uppsala, Sweden}
\affil{$^2$ Department of Physics, Royal Military College of Canada, PO Box 17000, Station Forces, Kingston, Ontario K7K 7B4, Canada}
\begin{abstract}
%
%
Magnetic fields in cool stars are ubiquitous but can still be challenging to characterize due to their complexity and relatively low strength. The polarization signature amplitudes are proportional to the field strength, and current studies of cool star magnetic fields are using circular polarization only since linear polarization is even weaker. However, circular polarization is only sensitive to the line-of-sight component of the magnetic field, meaning that many structural features are not recovered or may be misinterpreted when only circular polarization is used for reconstruction of stellar magnetic topologies. Linear polarization, on the other hand, is sensitive to the transverse component of the magnetic field and would provide a more complete picture of the magnetic field topology if combined with circular polarization. We have identified the first cool target, the RS CVn star II~Peg, suitable for full Stokes vector analysis. Using current instrumentation, we have succeeded in systematically detecting its linear polarization signatures with a precision and rotational phase coverage sufficient for magnetic mapping. Here we present the very first temperature and magnetic field maps reconstructed for a cool star using all four Stokes parameter spectra.
\end{abstract}
%
%
%
%
%
\section{Observations of II~Peg}
II~Peg is a K2IV, RS CVn binary star with $T_{\rm eff}$=4750 K and $P_{\rm rot}$=6.72 days \citep{Berdyugina98}. It is known to be very active and that is why we included it in our pilot survey where we tried to detect linear polarization in a small sample of cool stars at a level sufficient for imaging \citep{Rosen13}. The survey was successful and we have managed to obtain two sets of observations of II~Peg with sufficient phase coverage for Zeeman Doppler imaging (ZDI). The first set contains 7 observations performed between 25 September - 1 October 2012 and the second set consists of 12 observations obtained during 15 June - 1 July 2013. All observations were made at the Canada-France-Hawaii telescope with the spectropolarimeter ESPaDOnS \citep{Donati03} which has a resolving power of about 65000 and a wavelength coverage 370-1050 nm. We were not able to detect any linear polarization signatures in individual spectral lines and therefore applied the multi-line technique least-squares deconvolution (LSD) \citep{Donati97} to increase the S/N. In order for LSD to be applicable, all lines are assumed to be scaled versions of some mean profile.

\section{Zeeman Doppler imaging with LSD profiles}
The traditional approach of using LSD profiles for ZDI is to assume the observed LSD profiles behave like a real spectral line. The LSD profile is assigned some mean line parameters and is thereafter treated as a single line with those parameters. Local single line profiles are calculated using these mean line parameters, and are then integrated over the stellar disk. The disk-integrated profiles are then compared to the observed LSD profiles. 

A conceptual weakness of this approach concerns the fact that an LSD profile is an average over thousands of spectral lines. In general, its behavior cannot be accurately reproduced with a single spectral line as demonstrated by \citet{Kochukhov10}. In their study they showed that the single-line approximation requires fields below about 2~kG for Stokes $IV$ and that it can not reproduce Stokes $QU$ profiles at all.

Since we have Stokes $QU$ observations we are using a new approach to LSD profile modeling \citep{Kochukhov14}. A grid of synthetic LSD profiles is calculated from full polarized synthetic spectra covering the entire ESPaDOnS wavelength range where each profile corresponds to a different temperature, magnetic field strength, limb angle and magnetic field vector orientation. The local LSD profiles are obtained by an interpolation in this grid. These local profiles are then integrated over the disk into single LSD profiles which are compared to the observed LSD profiles. 

No assumptions are made about the behavior of the LSD profiles and they are used only to compress information and compare with observations. Using this approach, all four Stokes parameters can be interpreted regardless of magnetic field strength.

\section{Preliminary results}

\begin{figure*}
\centering
\includegraphics[scale=0.5]{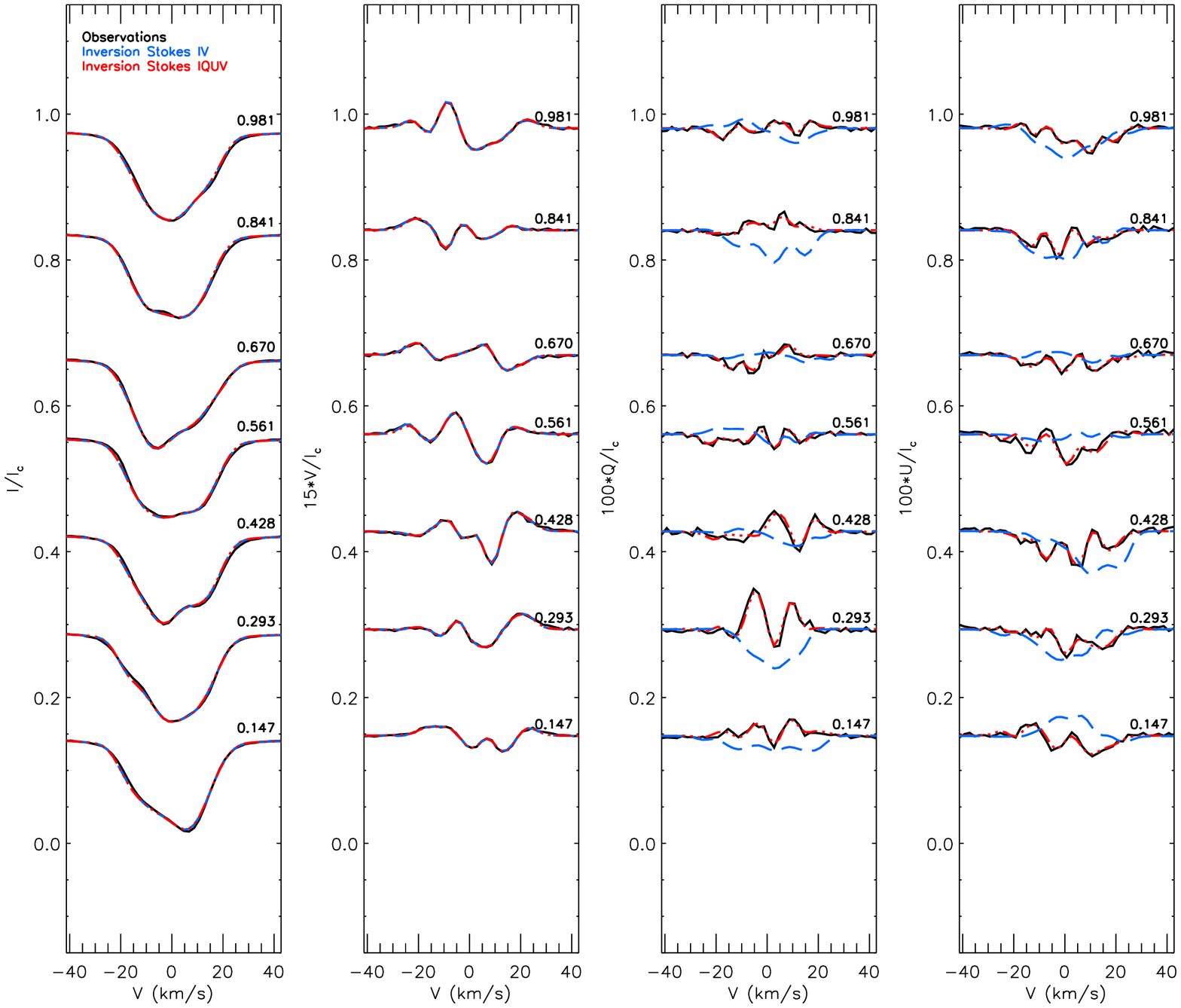} \\
\vspace{0.5cm}
\includegraphics[scale=0.47,angle=90]{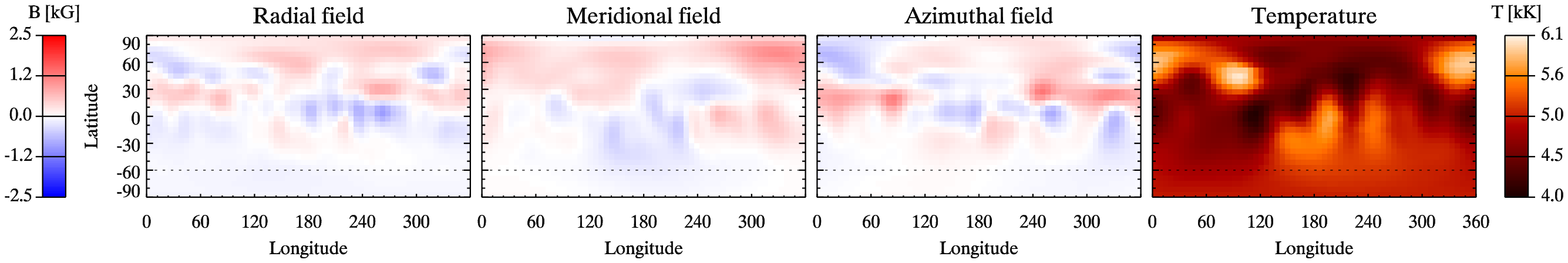} \\
\vspace{0.5cm}
\includegraphics[scale=0.47,angle=90]{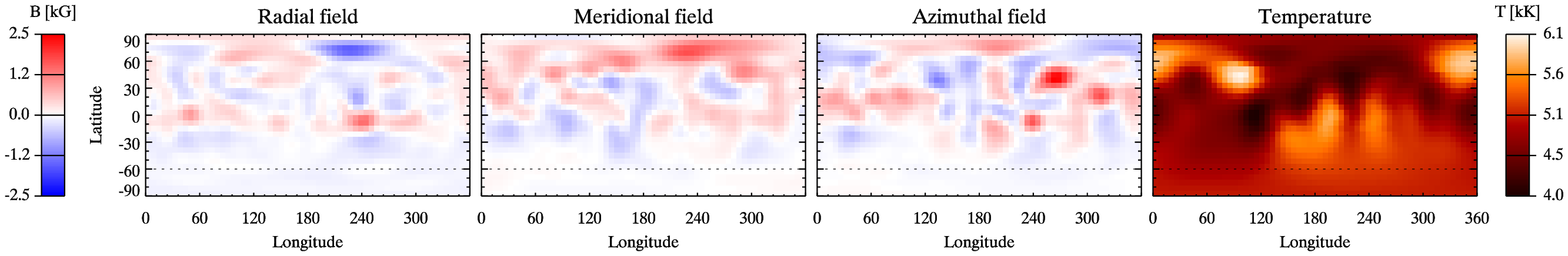}
\caption{Observations of II~Peg from September 25-October 1 2012. Line profiles of II~Peg where each of the four vertical panels show the LSD Stokes $IQUV$ profiles respectively. Profiles have been shifted vertically according to the rotational phase and the orbital radial velocity variation has been corrected. Stokes $Q$ and $U$ have been magnified by a factor of 100 while Stokes $V$ has been magnified by a factor of 15. The black solid lines represent the observations. The blue dashed line represents the model profiles when Stokes $IV$ was used in the inversion and the red dash-dotted lines represents the model profiles when Stokes $IQUV$ was used. The middle panel shows the reconstructed magnetic and temperature maps when only Stokes $IV$ was used in the inversion and the bottom panel shows the reconstructed magnetic and temperature maps when Stokes $IQUV$ was used in the inversion. 
}
\label{sep12_lines}
\end{figure*}

\begin{figure*}
\centering
\includegraphics[scale=0.5]{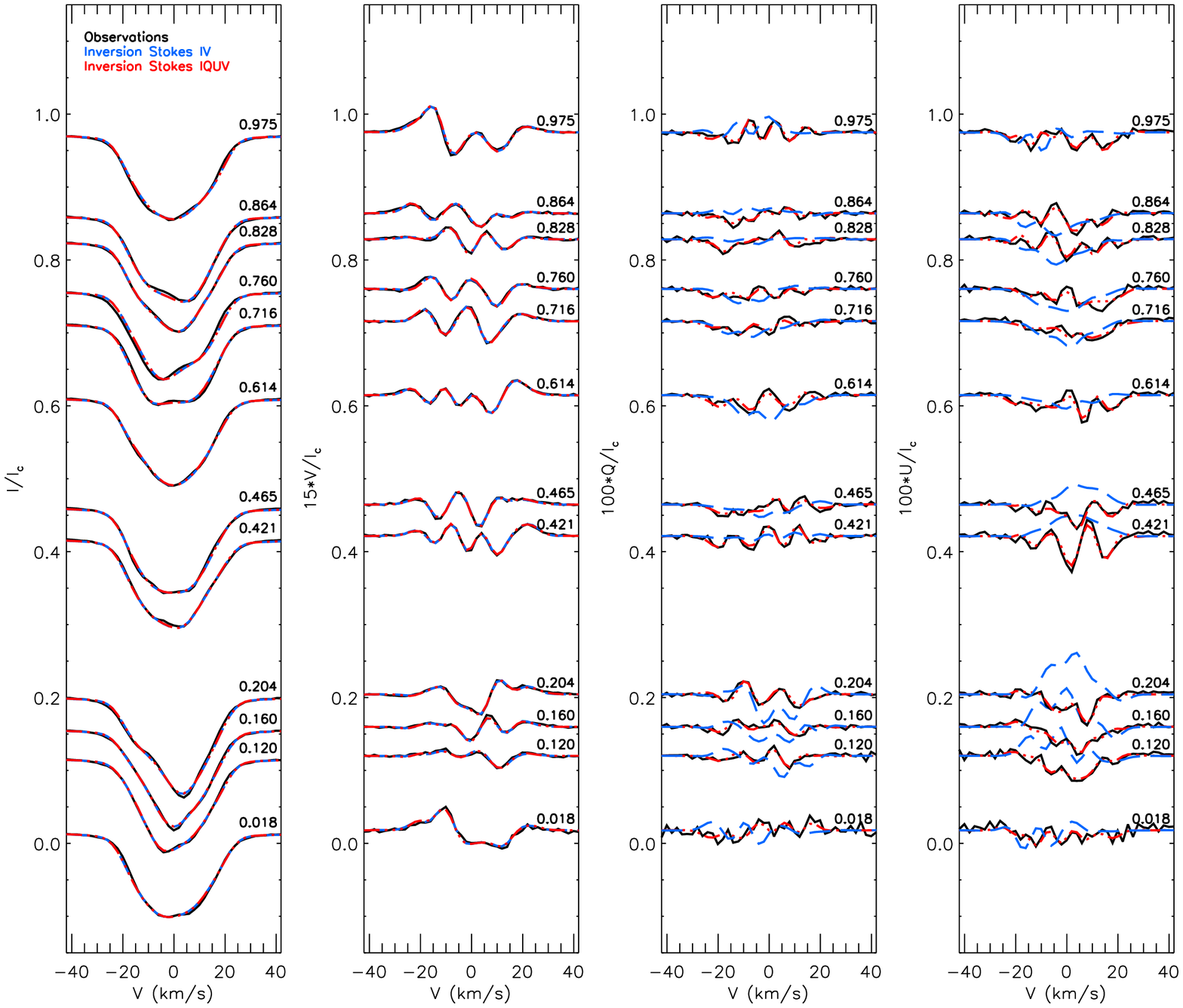} \\
\vspace{0.5cm}
\includegraphics[scale=0.47,angle=90]{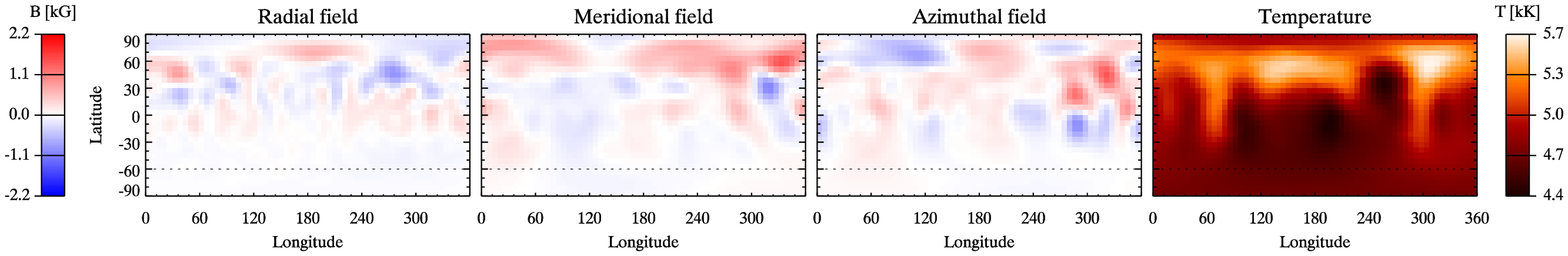} \\
\vspace{0.5cm}
\includegraphics[scale=0.47,angle=90]{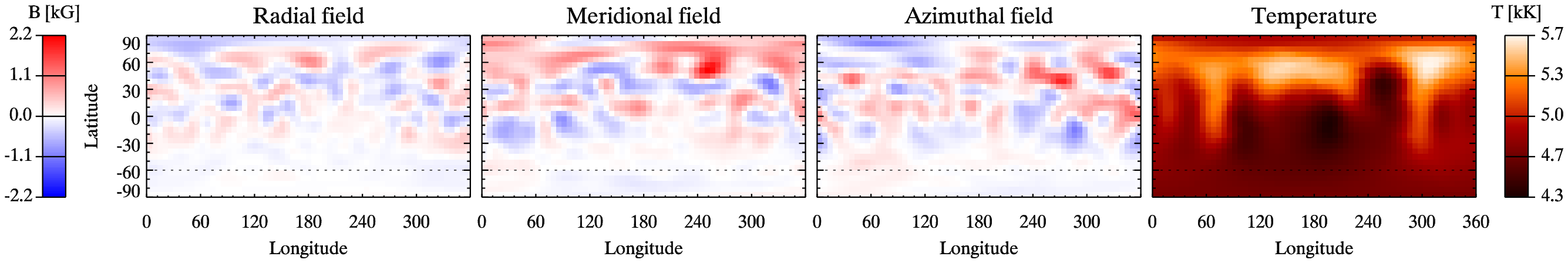}
\caption{Same as for Fig.~\ref{sep12_lines} but here the observations of II~Peg are from June 15-July 1 2013.  
}
\label{sum13_lines}
\end{figure*}

In Fig.~\ref{sep12_lines} and \ref{sum13_lines} the observed line profiles from the two observing periods are plotted together with the model profiles. These figures also show the corresponding reconstructed temperature and magnetic field maps using either Stokes $IV$ or Stokes $IQUV$ profile timeseries. 

The observed line profiles show that the activity seems to have been continuously high throughout both observing epochs. The difference in profile shape from one observing period to the other suggests the magnetic field has evolved during this time. When only Stokes $IV$ are used to reconstruct the magnetic field, the corresponding Stokes $QU$ model profiles of that magnetic map do not match the observed Stokes $QU$ profiles. However, when linear polarization is taken into account in the reconstruction process, the derived model profiles agree well with the observed Stokes $QU$ profiles. The influence of including Stokes $QU$ profiles in the inversion is reflected directly in the structure of the recovered map. When all four Stokes parameters are used, stronger features of the magnetic field become visible and the overall topology is more complex compared to when only Stokes $IV$ is used. This means that the same set of observed Stokes $V$ profiles can be fit by various different magnetic field distributions. It also suggests that Stokes $V$ alone can only, at best, give a rough overall picture of the complex magnetic field topologies of cool stars. 

%
%



%
%




%
%


\acknowledgments{
OK is a Royal Swedish Academy of Sciences Research Fellow, supported by the grants from Knut and Alice Wallenberg Foundation and Swedish Research Council. GAW is supported by a Discovery Grant from the Natural Science and Engineering Research Council of Canada (NSERC). The computations presented in this paper were performed on resources provided by SNIC through Uppsala Multidisciplinary Center for Advanced Computational Science (UPPMAX) under project snic2013-11-24.
}


\normalsize

\begin{references}

%
%
%
%
%

\bibitem[Berdyugina et al.(1998)]{Berdyugina98} Berdyugina, S.~V., Jankov, S., Ilyin, I., Tuominen, I., \& Fekel, F.~C.\ 1998, \aap, 334, 863 
\bibitem[Donati(2003)]{Donati03} Donati, J.-F.\ 2003, Solar Polarization, 307, 41
\bibitem[Donati et al.(1997)]{Donati97} Donati, J.-F., Semel, M., Carter, B.~D., Rees, D.~E., \& Collier Cameron, A.\ 1997, \mnras, 291, 658
\bibitem[Kochukhov et al.(2014)]{Kochukhov14} Kochukhov, O., L{\"u}ftinger, T., Neiner, C., Alecian, E., \& MiMeS Collaboration 2014, \aap, 565, A83 
\bibitem[Kochukhov et al.(2010)]{Kochukhov10} Kochukhov, O., Makaganiuk, V., \& Piskunov, N.\ 2010, \aap, 524, A5 
\bibitem[Ros{\'e}n et al.(2013)]{Rosen13} Ros{\'e}n, L., Kochukhov, O., \& Wade, G.~A.\ 2013, \mnras, 436, L10 

\end{references}
\end{document}